# Topological charge switching in trapped polariton condensates

Jaewon Kim[1], Hyun Gyu Song[2], Daegwang Choi[3,*], Yong-Hoon Cho[1,*]

[1]Department of Physics, Korea Advanced Institute of Science and Technology (KAIST), Daejeon 34141, Republic of Korea
[2]Nanophotonic System Research Center, Korea Institute of Science and Technology (KIST), Seoul 02792, Republic of Korea
[3]Department of Physics and Semiconductor Science, Gachon University, Seongnam 13120, Republic of Korea
*e-mail: yonghcho@kaist.ac.kr, dgchoi@gachon.ac.kr

**Topological charges in photonic systems provide a robust degree of freedom with direct applications in optical information processing. Bound states in the continuum (BICs) in photonic crystal slabs inherently carry such quantized topological charges, yet they are topologically protected, making active reconfiguration fundamentally challenging. Here, we demonstrate topological charge switching in trapped BIC polariton condensates in a photonic crystal, where spatial modulation of the photonic crystals supports multiple confined states, each carrying a distinct charge. Trapped condensates within the Dirac bandgap are expelled into the leaky band via polariton blueshift, enabling single-mode selection by solely tuning the excitation power. This work offers a platform for optically controlling topological charges by harnessing polariton nonlinearity.**

## Main

Vortex beams carrying topological charges have established themselves as a key approach in photonics, enabling optical communication [1–3], optical tweezers [4,5], and quantum information [6,7]. To generate such beams in a compact form, vortex lasers based on nanophotonic platforms such as microring resonators, microcavities, and metasurfaces have been actively studied [8–10]. In particular, ultracompact photonic slab structures such as metasurfaces have attracted growing interest, as they naturally support bound states in the continuum (BICs), non-radiating resonances with theoretically infinite quality factors [11–13], whose intrinsic topological charges manifest as polarization vortices in momentum space [14,15]. These properties have enabled single-mode, directional BIC lasers with high quality factors across a wide range of material systems [16–21].

Actively tuning the topological charge within a single device would enable more versatile applications, yet this remains a considerable challenge. Although approaches such as engineered optical pumping schemes, microelectromechanical system (MEMS)-based actuation, and temperature modulation have been reported [10,22–25], they rely on bulky and slow approaches that are difficult to implement in practical applications. Moreover, since the topological charges of BICs are topologically protected and determined by the structural parameters of the fabricated device, they are fundamentally difficult to reconfigure dynamically. A key challenge, therefore, is to realize a switching mechanism between topological states with a simple control parameter, while maintaining single-mode operation.

Exciton–polaritons in a photonic crystal, formed by strong coupling between excitons and BIC modes [16,26], however, offer an inherent tunability through interaction-induced nonlinearity that enables active control of the topological states. Additionally, the spatial potential implemented by lattice constant modulation supports discrete polariton states, each carrying a distinct topological charge. Here, we demonstrate all-optical switching of topological charge in trapped polariton condensates, realized within a photonic crystal well structure. We first characterize the polariton band dispersion of the photonic crystal well and confirm quantum confinement of the polariton states as a function of well width. We then demonstrate polariton condensation into individual trapped states and characterize their polarization vortices, revealing topological charges of $q$ = +1 to +4. Finally, we realize all-optical topological charge switching driven solely by tuning the excitation power. The switching is enabled by interaction-induced polariton blueshift in combination with the Dirac bandgap, acting as an intrinsic spectral filter that suppresses condensation in states entering the barrier continuum and thereby maintains single-mode condensation.

## Results

### Photonic crystal well structures for quantized polariton states.

We consider a GaAs/$Al_{0.15}Ga_{0.85}As$ multi quantum well (MQW) on an AlAs cladding layer, patterned into a one-dimensional (1D) photonic crystal structure, shown in Fig. 1a. In this slab structure, the guided mode resonances (GMRs) form Dirac-like band dispersions with a symmetry-protected BIC at the Γ point [11,27] (see Supplementary Section S1 for details). These modes are strongly coupled to the excitonic resonance of the MQW, forming polariton states [16]. The energy of these polariton modes is set by the structural geometry, with a larger lattice constant shifting the photonic modes to lower energies. Therefore, spatial modulation of the lattice constant engineers a photonic potential analogous to a finite quantum well in the Dirac equation, allowing polaritons to experience a spatially varying potential landscape. This type of local structure modulation has also been explored in conventional distributed Bragg reflector (DBR)-based polariton systems through a variety of methods to apply polariton potential [28–32]. Since the BIC polariton state is located at the lower Dirac band with a negative effective mass, a local potential maximum acts as an effective trap for spatial confinement [33,34]. Therefore, we design the well region with a smaller lattice constant and the barrier region with a larger lattice constant, where the resulting energy offset of the Dirac band in the well region creates the lateral confinement potential for the BIC polaritons (Fig. 1c). The confined polariton states trapped in the photonic crystal well are quantized in energy, forming discrete modes within the gap between the lower and upper Dirac bands of the barrier. As the excitation power increases, the trapped condensate blueshifts due to polariton interaction-induced nonlinearity. Since the blueshift of the gap-confined mode exceeds that of the barrier Dirac bands, the $n = 1$ state eventually reaches the upper Dirac band of the barrier and leaks into the barrier continuum, while the $n = 2$ state shifts into the gap, resulting in an abrupt switching of the condensate mode, as illustrated in Fig. 1d and 1e. This mechanism underlies the topological charge switching demonstrated in this work, as discussed in detail below.

To experimentally realize and characterize the proposed topological charge switching, we first perform angle-resolved reflectance spectroscopy on a uniform photonic crystal structure to characterize the photonic band dispersion for the transverse-electric (TE) mode. We employ a well lattice constant of $a = 240$ nm and a barrier lattice constant of $a = 244$ nm, with the corresponding angle-resolved reflectance spectra shown in Fig. 1b. The dispersion of the well structure is clearly gained energy offset by ~ 20 meV relative to the barrier structure, providing the photonic potential for lateral confinement. From the measured spectra, the heavy hole exciton resonance is identified at 1.553 eV, with an extracted Rabi splitting of 15 meV (see Supplementary Section S2 for details).

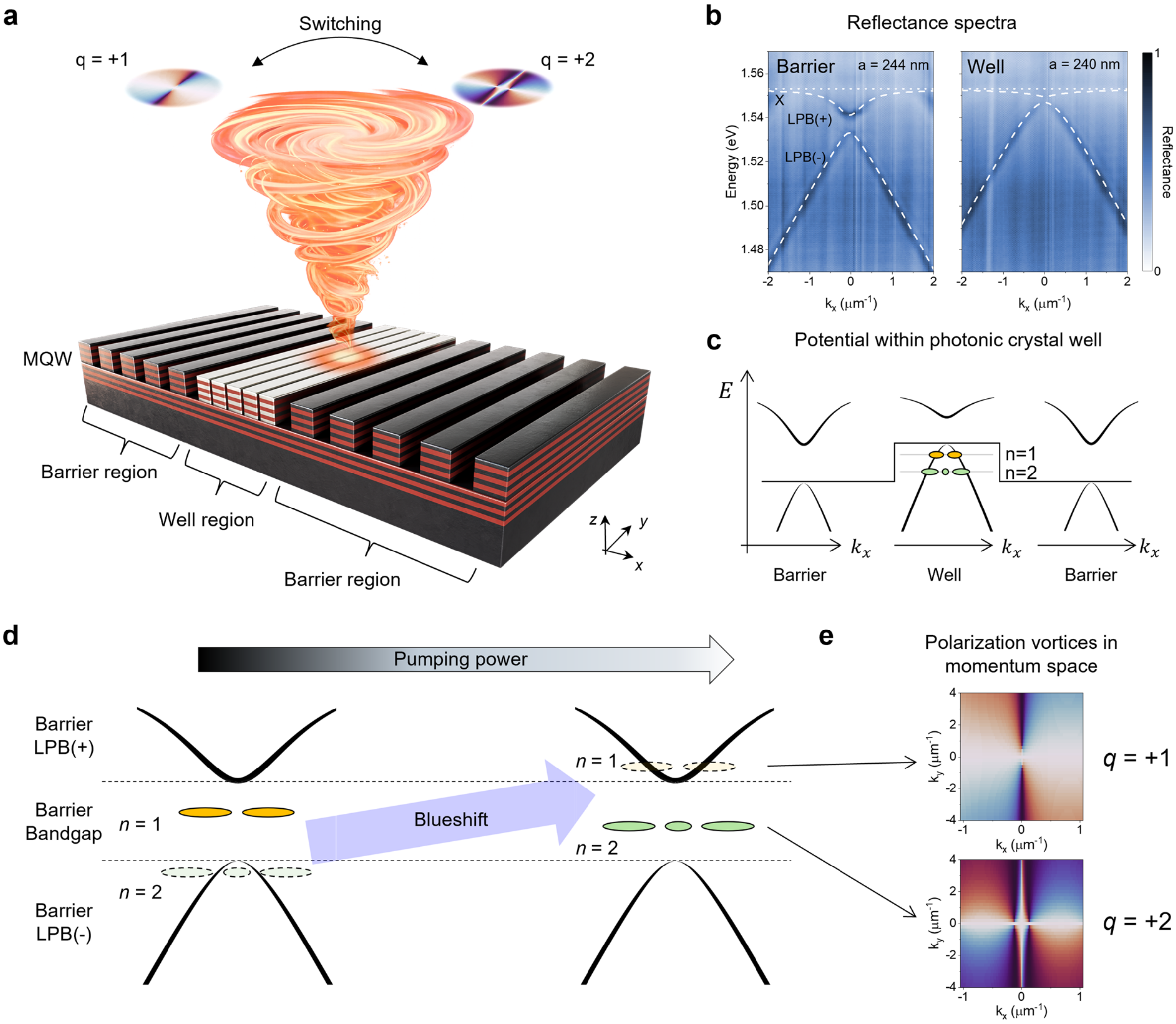

**Fig. 1. Exciton-polariton from photonic crystal well. a**, Schematic illustration of an exciton–polariton confined in a photonic crystal well formed by locally modifying the lattice constant of a 1D photonic crystal. **b**, Polariton dispersion of barrier ($a = 244$ nm) and well lattice ($a = 240$ nm) measured by angle-resolved reflectance. LPB (+) and LPB (-) denote the symmetric and antisymmetric lower polariton branches, corresponding to the upper and lower Dirac bands, respectively. **c**, Confinement mechanism for polaritons with negative effective mass. The barrier region opens a photonic bandgap that spatially confines the polariton modes in the well region. **d**, Topological switching mechanism of confined polariton condensate in photonic crystal well. Increasing the pump power induces a blueshift from polariton-polariton interaction, which drives a transition between condensate states with different topological charges. **e**, Simulated momentum space polarization vortex with a topological charge of $q$ = +1 and $q$ = +2. Each vortex corresponds to the confined ground state and first excited state, respectively.

## Width Modulation

To confirm the quantum confinement of the photonic crystal well, we perform angle-resolved photoluminescence (PL) spectroscopy as a function of well width under continuous wave (cw) non-resonant excitation below the condensation threshold. At a well width of $w$ = 2.88 μm, a ground state ($n$ = 1) emerges within the Dirac bandgap region, shown in Fig. 2a. As the well width $w$ is increased to 9.60 μm and 19.2 μm, successive excited states $n$ = 2 and $n$ = 3 and 4 appear within the gap, respectively (Figs. 2b and 2c). These confined states are also visible in angle-resolved reflectance measurements, confirming confinement arises solely from structural modulation (see Supplementary Section S3 for details).

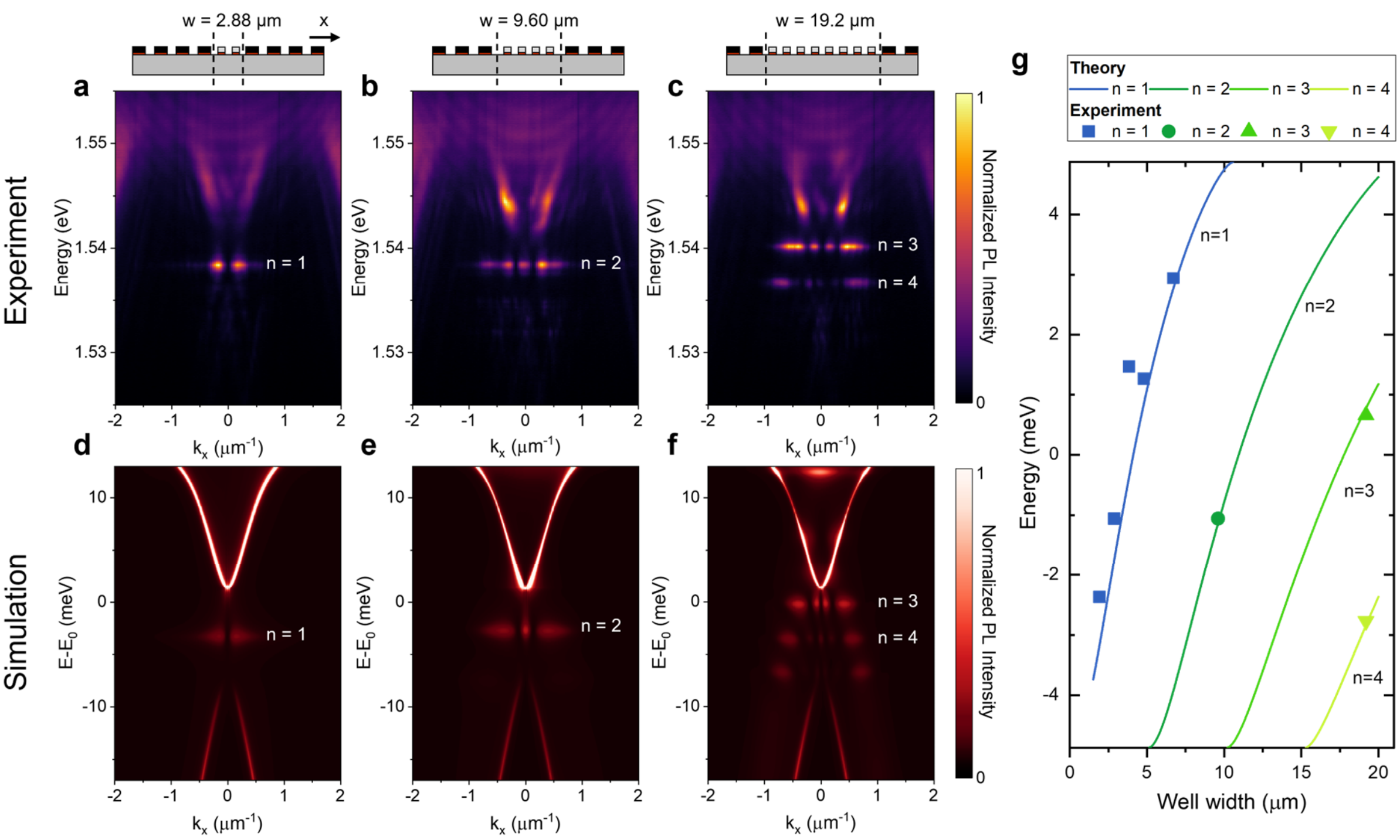

**Fig. 2. Discretized confined states of polaritons. a-c,** Experimental dispersion of confined polaritons with a well width of $w$ = 2.88, 9.60, and 19.2 μm under the excitation below the condensation threshold. Discretized confined polariton mode with quantum number of $n$ = 1, $n$ = 2, and $n$ = 3, 4 appear, respectively, on the width. **d-f**, Simulated dispersion of confined polaritons with the same set of well widths as in a-c. **g**, Comparison between confined state energy obtained from theoretical calculation (solid line) and those extracted from experimental results (filled symbols).

These photonic crystal polaritons can be modeled by a 1D non-Hermitian Dirac Hamiltonian of the two counter-propagating GMR and exciton resonance basis [26,35].

$$H_{ep} = \begin{pmatrix} E_x - i\gamma_{nr} & 0 & g & 0 \\ 0 & E_x - i\gamma_{nr} & 0 & g \\ g & 0 & -ihv_g\partial_x - i\gamma + V_p(x) & U + i\gamma \\ 0 & g & U + i\gamma & ihv_g\partial_x - i\gamma + V_p(x) \end{pmatrix} \quad (1)$$

where $E_x$ is the exciton resonance energy, $\gamma_{nr}$ is the non-radiative decay rate of the exciton, $g$ is the exciton-cavity coupling parameter, $v_g$ is the group velocity of the GMR, $\gamma$ is the radiative loss of the GMR, $U$ is the diffractive coupling strength of the counter-propagating wave, and $V_p(x)$ is the photonic potential. Steady-state solution of the Hamiltonian is simulated by Monte-Carlo sampling realizations from different random initial conditions (see Methods for detailed simulation method). Numerical simulations of the far-field dispersion for $w = 2.88$, 9.60, and 19.2 μm are shown in Figs. 2d–2f, in excellent agreement with the experimental results.

We also conduct analytic calculations to derive the eigen-energies of the confined states as a function of well width. (see Supplementary Section S4 for details). The analytically derived eigen-energies show good agreement with the experimental results (Fig. 2g). Notably, when the energy of a confined state exceeds the barrier bandgap, no bound state solution exists, as the mode leaks into the barrier continuum.

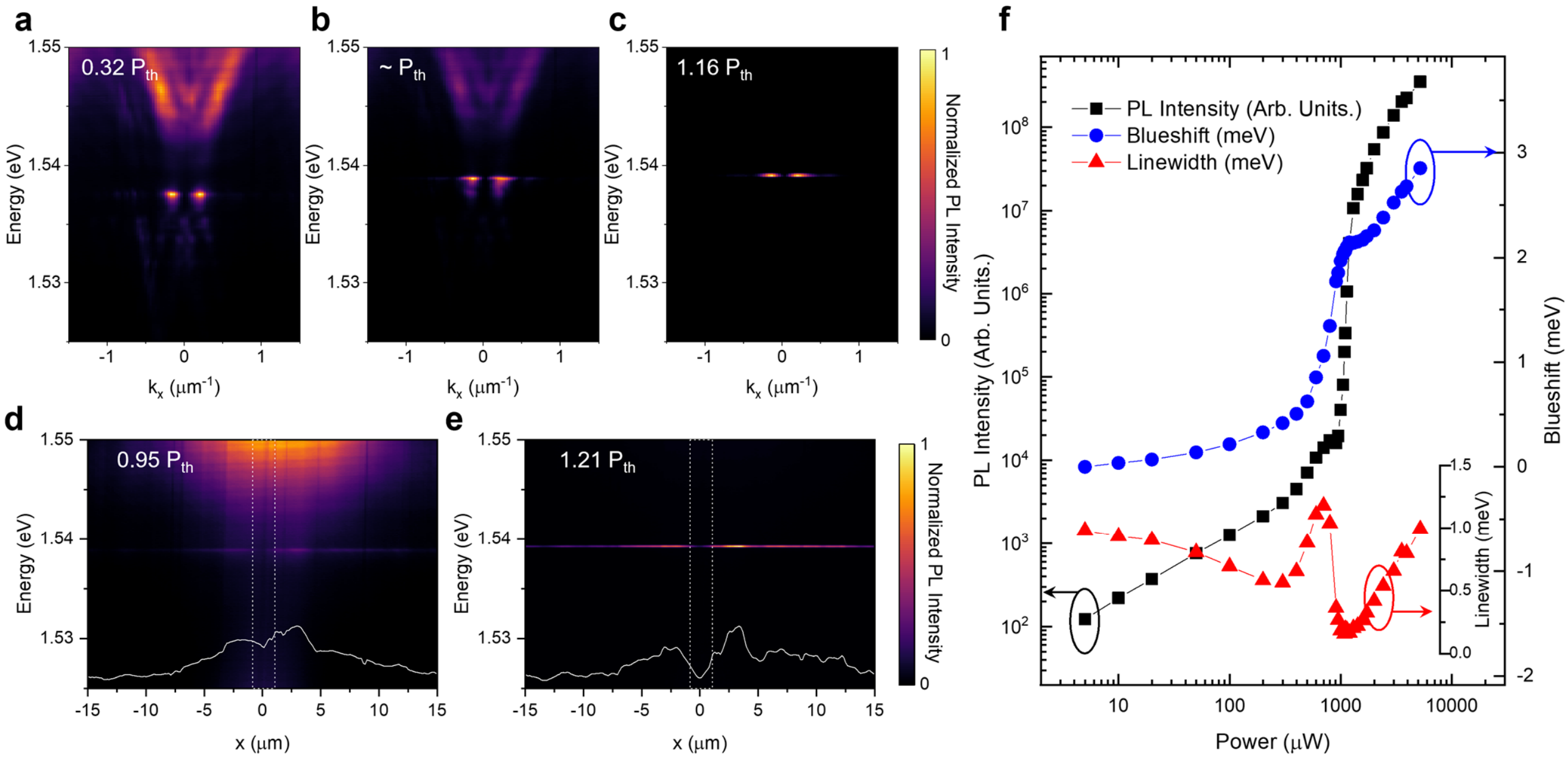

**Fig. 3. Polariton condensation into a confined mode. a-c,** Angle-resolved PL spectrum with an excitation power of $P = 0.32\ P_{th}$, $P \sim P_{th}$ and $P = 1.16\ P_{th}$ for a well width of $w = 1.92$ μm. **d, e,** Spatial-resolved PL spectrum before and after the condensation threshold. Dashed white lines indicate the region of the well. **f,** Integrated PL intensity (Black), blueshift (Blue), and linewidth (Red) of the confined state by varying excitation power. Superlinear increase of the PL intensity and spectral narrowing indicate the polariton condensation. The blueshift of the polariton mode arises from interaction-induced energy renormalization, with the dominant interaction changing from polariton-exciton reservoir interactions below threshold to polariton-polariton interactions above threshold.

**Polariton condensation of trapped modes**

We perform angle-resolved PL spectroscopy to investigate condensation behavior of these confined polariton states using pulsed excitation. For a well width of $w = 1.92$ μm, angle-resolved PL spectra below, near, and above the condensation threshold are shown in Figs. 3a–3c. Both barrier continuum dispersion and the confined $n = 1$ state are shown below the threshold in Fig. 3a. Above the threshold of $P_{th} = 16.8$ μJ/cm$^2$, polaritons condense to the trapped state in Fig. 3c. To confirm confinement of the polaritons, spatial-resolved PL measurement is preformed, shown in Figs. 3d and 3e. Below the condensation threshold, an exciton reservoir forms at the laser spot (Gaussian spot with a full width at half maximum of 7.6 μm) and feeds polaritons into both the barrier band and the confined ground state. Above the condensation threshold, polaritons strongly accumulate in the confined ground state. Power-dependent PL intensity, energy, and linewidth of the trapped state for $w = 1.92$ μm are shown in Fig. 3f. The PL intensity increases superlinearly above the condensation threshold, accompanied by a linewidth narrowing due to the onset of coherence, followed by broadening above threshold from polariton–polariton interactions. Similar single-mode polariton condensation behaviors and spatial confinement on $n = 2$, and 3 are observed in samples with larger well width (see Supplementary Section S5 for details). Furthermore, phase-resolved interferometry reveals the spatial phase distribution of the condensate depending on the well width, with the number of π-phase jumps matching the quantum number $n$ (see Supplementary Section S6 for details).

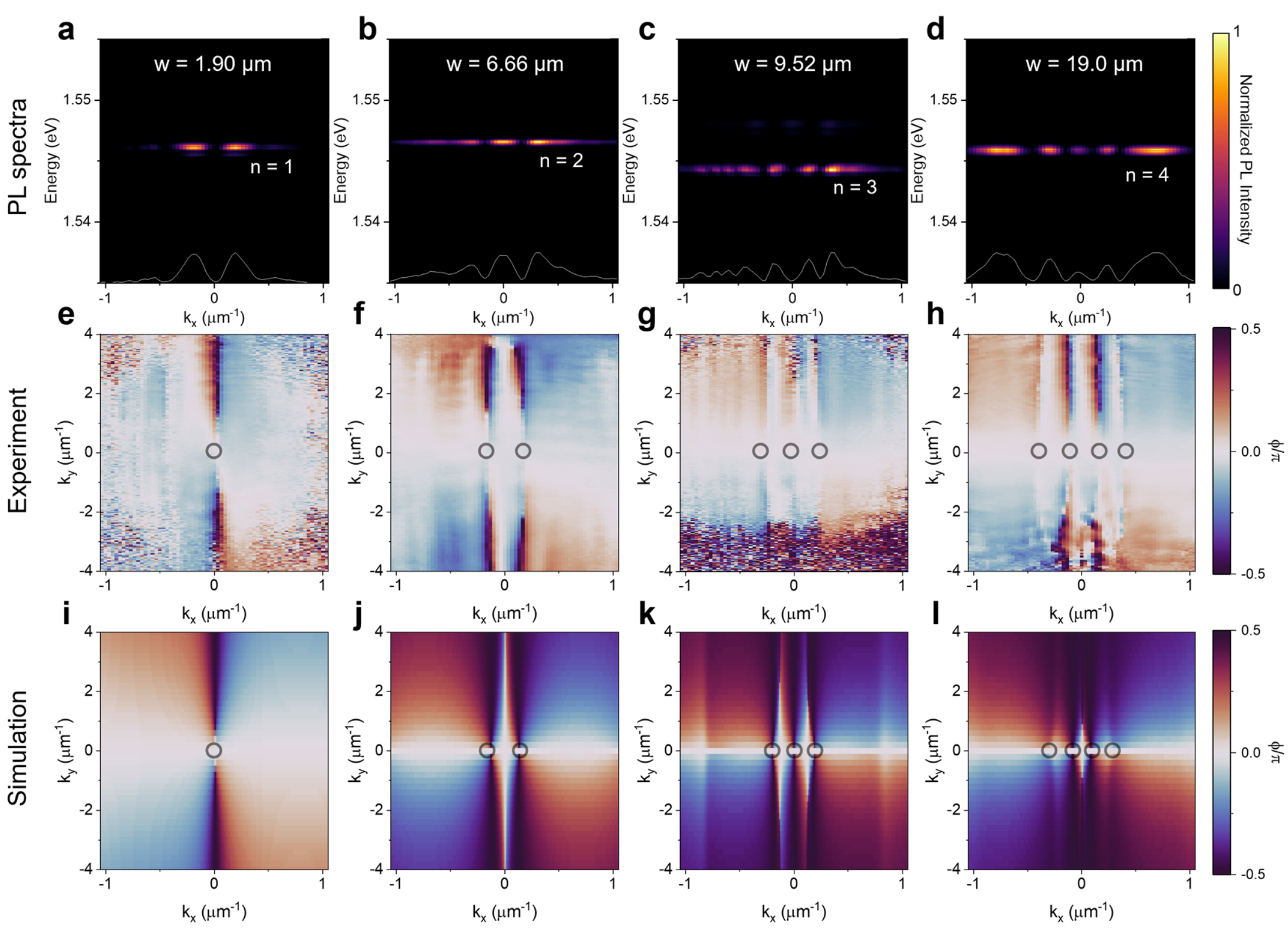

**Fig. 4. Polarization vortex from a single confined mode condensate. a-d,** Polariton condensate dispersion for samples with various well widths $w$ = 1.90, 6.66, 9.52, and 19.0 μm under the excitation above the condensation threshold. Each sample exhibits a single-mode condensation in a confined state with quantum numbers of $n$ = 1, 2, 3, and 4, respectively. **e-h,** Experimental polarization vortices from polarization- and angle-resolved PL images. Open circles indicate the polarization singularities. **i-l,** Simulated polarization vortices reconstructed from the Dirac spinor components of the corresponding confined modes.

**Topological polarization vortex of confined polariton condensate**

We investigate topological properties of the trapped polariton condensate using polarization-resolved measurements, where the topological charge is defined by the winding number of the polarization vector field [14]. In a uniform photonic crystal, the BIC polariton condensate carries a topological charge of $q$ = +1 in momentum space, as characterized in Supplementary Section S8. For the trapped polaritons, the polarization characteristics are not solely determined by the topology inherited from the BIC. Figures 4a–4d show angle-resolved PL spectra of trapped polariton condensates above the condensation threshold for well widths of $w$ = 1.90, 6.66, 9.52, and 19.0 μm. In each case, condensation occurs into a single confined state, with the condensate exhibiting a distinct momentum-space envelope function. This confined envelope function modulates the polarization topology, giving rise to trapped states with distinct topological charges [26]. A polarization vortex in momentum space is captured from polarization-resolved PL measurement (see Method for detailed information). Polarization-resolved momentum-space imaging reveals well-defined polarization vortices for each condensate state, carrying topological charges of $q$ = +1 to +4 for $n$ = 1 to 4, respectively (Figs. 4e–4h). For the simulation, polarization vortices of each trapped state are derived from a Dirac spinor component, as a result of Monte-Carlo sampling of the governing Hamiltonian Eq. (1). Simulations of each trapped state polarization vortex are shown in Figs. 4i–4l, closely matching the experimental results.

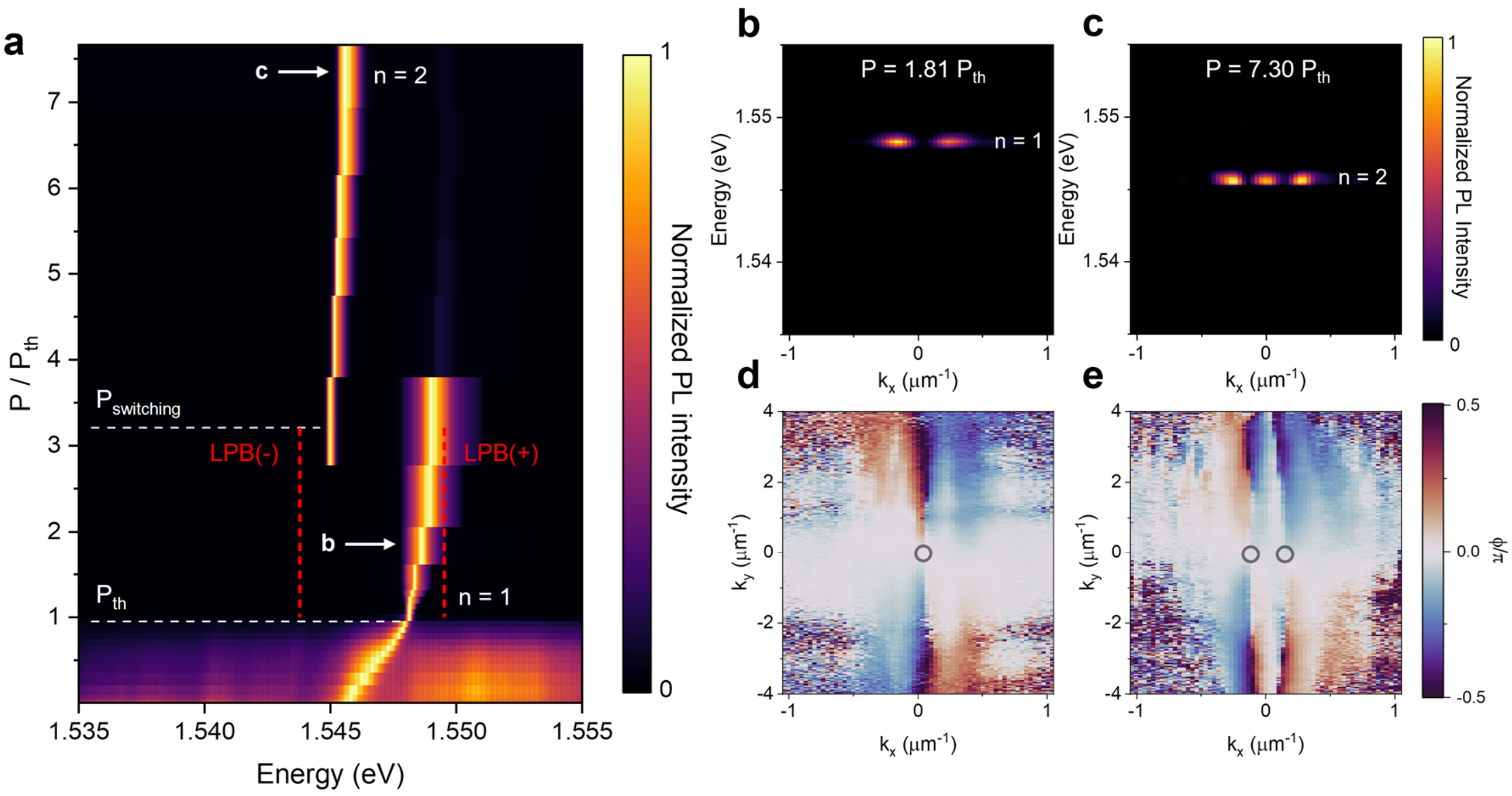

**Fig. 5. Topological switching between confined mode condensates. a,** Excitation power-dependent PL spectrum of a confined polariton sample with well width of $w$ = 2.86 μm. Red dashed lines indicate the barrier band edges, Red dashed lines indicate the barrier band edges, measured below the condensation threshold using a sample with only the barrier lattice. The low power region is magnified to highlight condensation to the $n$ = 1 mode. With increasing excitation power, the condensate blueshifts due to the polariton-polariton interaction, and the $n$ = 1 mode merges into the barrier continuum, inducing a transition from $n$ = 1 to $n$ = 2 condensate. **b, c,** Angle-resolved PL spectrum at $P$ = 1.81 $P_{\text{th}}$ and $P$ = 7.30 $P_{\text{th}}$, showing $n$ = 1 and $n$ = 2 condensate, respectively. **d, e,** Measured topological polarization vortices corresponding to b and c, revealing switching between topological condensates.

## Switching between topological polariton condensates

We now demonstrate switching behavior between these distinct topological charges within a single photonic crystal well structure. We use a sample with a well width of $w$ = 2.86 μm, which hosts a confined state of $n$ = 1 in close proximity to the upper Dirac band of the barrier. Three distinct regimes of polariton behavior are observed as a function of excitation power, as described in the power-dependent PL spectrum in Fig. 5a. Below the condensation threshold, polaritons exhibit a broad spectrum comprising both the barrier band and the confined state. Above the threshold, polaritons condense into the $n$ = 1 state, yielding a single-mode condensate as shown in Fig. 5b. In this regime, the polariton condensate carries a topological charge of $q$ = +1, as confirmed in Fig. 5d. Upon further increasing the excitation power, polariton–polariton interactions blueshift the $n$ = 1 state until its energy reaches that of the upper Dirac band of the barrier, at which point it leaks out into the barrier. Simultaneously, the $n$ = 2 state blueshift into the bandgap, and polaritons condense into the $n$ = 2 state instead of the $n$ = 1 state. Following this second transition, single-mode condensation in the $n$ = 2 state is observed, as shown in Fig. 5c. Analysis of the k-space polarization profile confirms that this condensate carries a topological charge of $q$ = +2, as shown in Fig. 5e. These results clearly demonstrate that switching between topological condensates is achievable within a photonic crystal well, driven solely by excitation power without any structural reconfiguration. Unlike conventional polariton systems where increased pumping induces multi-mode coexistence with continuous blueshifts, our Dirac band-bounded configuration suppresses multimode states, enabling deterministic single-mode topological switching.

## Conclusion

In this work, we demonstrate topological switching of exciton-polariton condensates from a photonic crystal well. Notably, to the best of our knowledge, such topological switching driven solely by excitation power has not been previously reported, either in polariton condensation or in photonic lasing. This is made possible by three key features inherent to the exciton-polariton system in a photonic crystal well. (i) The photonic crystal well provides quantized states with well-defined topological charges inherited from the BIC. (ii) The exciton-polaritons exhibit nonlinear polariton–polariton interactions that enable active tuning of the polariton energy. (iii) The barrier band of the photonic crystal well acts as a spectral filter that enforces single-mode condensation, as polaritons lying outside the barrier bandgap leak into the surrounding barrier and are suppressed.

While multistate polariton condensation has been widely studied in conventional DBR-based systems [36–39], mode switching in such structures is challenging as excitation power increases, multiple polariton modes blueshift simultaneously without disappearing, making it difficult to maintain single-mode occupancy. Our photonic crystal well structure overcomes this through two key features. First, the upper Dirac band is inherently leaky, such that trapped states that reach this band leak into the radiative continuum, preventing multi-mode occupancy. Second, spatial trapping enhances polariton–polariton interactions relative to the delocalized Dirac band modes, causing trapped states to blueshift faster than the Dirac bands (see Supplementary Section S7 for details). Together, these two features — the leaky upper Dirac band and the inherently strong nonlinearity of the trapped polaritons — are what make robust topological mode switching possible in this system.

Previous photonic crystal laser studies have mainly used structurally defined or pump-induced potentials to localize modes and optimize Q-factor, mode volume, and lasing threshold [26,40–45]. In contrast, our system combines a passive photonic crystal potential with excitation-dependent polariton nonlinearity, using the excitation power as an additional dynamic control knob. This enables topological switching without modifying the structure or pump profile.

Beyond the photonic crystal well, this platform offers a versatile route to engineer arbitrary potential landscapes through structural modulation, extending the rich tradition of polariton potential engineering from DBR-based systems into the photonic crystal regime. This opens the possibility of realizing polariton molecules [46] and 1D polariton lattices [47–49], with further extension to two-dimensional lattice geometries [50–54].

Regarding the switching itself, while the present work employs optical excitation, the ultrafast polariton lifetime provides a natural pathway toward high-speed optoelectronic devices. Temporal modulation of the excitation laser could enable switching speeds in the GHz regime, limited only by the polariton lifetime and relaxation time. Furthermore, extending this platform to room-temperature material systems such as perovskites, organic materials, and wide-bandgap semiconductors would bring topological charge switching closer to practical applications in optical communications and information processing.

**Acknowledgements**
This work was supported by the National Research Foundation of Korea (NRF) grant (RS-2023-00284018, RS-2025-25446463, RS-2024-00416538), by the Institute of Information & Communications Technology Planning & Evaluation (IITP) grant (RS-2024-00397959, RS-2025-25464990) funded by the Korea government (MSIT). This work was supported by the Gachon University research fund of 2026 (GCU-202601470001). This work was supported by the National Research Foundation of Korea (NRF) grant funded by the Korea government (MSIT) (RS-2026-25492581). The authors gratefully acknowledge Hyungsoon Choi and Suhong Choi for their valuable support in establishing the experimental setup.

**Author Contributions**
J.K., D.C., and Y.-H.C. initiated this work. J.K., D.G., and H.-G.S. designed and conceptualized this work. J.K. developed the sample design and carried out numerical simulations and analytic modeling. J.K. and D.C. fabricated the samples. J.K. conducted optical experiments and data collection. J.K. and D.C. performed all data analysis and visualization. J.K. and D.C. wrote the manuscript with input from all authors. The project was supervised by D.C. and Y.-H.C. All authors reviewed and revised the manuscript.

**Competing interests statement**
The authors declare no competing interests.

## Methods

### Sample fabrication

First, a multi-quantum well structure is grown by metal-organic chemical vapor deposition (MOCVD) [55]. A 500 nm AlAs layer is grown on a GaAs substrate. This is followed by the growth of alternating 30 nm $Al_{0.15}Ga_{0.85}As$ layers and 10 nm GaAs layers for 16 pairs, forming a multi-quantum well structure. Low-temperature ($T$ = 4 K) micro-photoluminescence is used to confirm exciton resonance. To design a photonic structure where the BIC mode is resonant with excitons, we utilize a 1D grating photonic crystal. The design of the photonic structure is performed using rigorous coupled-wave analysis (RCWA) from Stanford S4 [56]. In the simulation, by injecting plane waves over a range of angles and energies and measuring the reflectance, we simulate the photonic dispersion coupled to the far field. Realization of the structure is performed by electron beam lithography and ICP-RIE dry etching [57]. In the photonic crystal well sample, the barrier lattice extends 150 unit cells on both sides of the well. The structure is conformally deposited with 8 nm of thin $Al_2O_3$ by atomic layer deposition, which passivates surface defects from etching and improves polariton linewidth.

### Simulation

We perform numerical simulations based on the time evolution of an effective exciton–polariton Hamiltonian to calculate the dispersion and spatial mode profiles of the polariton from a photonic crystal well. The system is described by the exciton–polariton Hamiltonian $H_{\mathrm{ep}}$ in equation (1). The polariton spinor wavefunction evolves according to $\partial_t\Psi = -(i/\hbar)H_{\mathrm{ep}}\Psi$. We use a Gaussian-enveloped random initial condition to excite modes spatially localized in the well. The time evolution is calculated using a split-step Fourier method combined with a discrete time domain propagation scheme, where the kinetic evolution is performed in momentum space and the potential evolution is performed in real space. After the wavefunction is stabilized into the eigenmodes, the simulated field is Fourier-transformed to obtain the momentum-resolved dispersion. Note that considering interference of the photonic part of the Dirac spinor, we can simulate the far-field dispersion or spatial distribution. The polarization vortex of the BIC mode is also simulated by extracting the far-field momentum-space polarization texture from the relative amplitude and phase of the Dirac spinor components obtained through the Hamiltonian time-evolution calculations.

### Measurements

Angle-resolved and spatial-resolved measurements are performed using a 50× objective lens with numerical aperture 0.55 for both excitation focusing and signal collection. The spectra and images are measured with a spectrometer (Princeton Instruments) and a charged-coupled device camera (Princeton Instruments). Reflectance measurements are conducted using a broadband halogen lamp. PL measurements are performed using a femtosecond Ti:sapphire laser, which is operated in either continuous-wave or 80 MHz pulsed mode depending on the mode-locking condition. Polarization-resolved PL measurements are conducted by placing a half-wave plate and a linear polarizer in the collection path. The PL intensity is resolved in different linear polarization bases, and the polarization angle is reconstructed from the polarization-dependent intensity distributions to extract the polarization vortex structure. Interference measurements are performed using a Mach-Zehnder-type interferometric setup. The PL signal from the sample is divided into two paths by a beam splitter. In one arm, the signal is transmitted through a pinhole at the back focal plane to generate a Gaussian reference beam, while in the other arm, the real-space PL image is passed through a delay line mounted on a computer-controlled translation stage (Thorlabs). The two beams are recombined on a charge-coupled device (CCD) camera to interfere the real-space image with the Gaussian reference. A small relative tilt between the reference beam and the PL image introduced a finite carrier frequency, allowing the interference signal to be separated from the background intensity components.

**Data availability**

The datasets used in this study are available from the corresponding author upon reasonable request.

# Supplementary information: Topological charge switching in trapped polariton condensates

Jaewon Kim[1], Hyun Gyu Song[2], Daegwang Choi[3,*], Yong-Hoon Cho[1,*]

[1]Department of Physics, Korea Advanced Institute of Science and Technology (KAIST), Daejeon 34141, Republic of Korea
[2]Nanophotonic System Research Center, Korea Institute of Science and Technology (KIST), Seoul 02792, Republic of Korea
[3]Department of Physics and Semiconductor Science, Gachon University, Seongnam 13120, Republic of Korea
*e-mail: yonghcho@kaist.ac.kr, dgchoi@gachon.ac.kr

## S1. PHOTONIC BOUND STATE IN THE CONTINUUM FROM A 1D PHOTONIC CRYSTAL SLAB

In the slab structure, guided modes form below the light cone and are therefore lossless, but cannot be extracted into free space. To enable far-field access, a one-dimensional (1D) photonic crystal is introduced by etching the waveguide surface. When the lattice constant satisfies a second-order Bragg condition, the guided modes couple to the radiative continuum above the light cone, allowing them to be accessed through conventional far-field optics.

The photonic crystal also provides diffractive coupling between the two counter-propagating guided modes. Because this coupling is non-Hermitian, the counter-propagating modes hybridize into two branches that are not only energy-split but also exhibit distinct non-Hermitian behavior: one hybridized mode (the leaky guided-mode resonance, leaky GMR) acquires twice the radiative loss, while the other (the bound state in the continuum, BIC) is ideally lossless. The energy of the bandgap center coincides with the second-order Bragg condition, which is set directly by the photonic crystal's lattice constant, following the relation of $E_c = \frac{hc}{n_{\mathrm{eff}}a}$. We used rigorous coupled-wave analysis (RCWA), implemented in the Stanford $S^4$ package, to design the photonic crystal structure [1]. The resulting photonic dispersion and electric-field distributions confirm BIC formation, described in Fig. S1.

To experimentally verify BIC formation, we performed angle-resolved reflectance measurements on the uniform-lattice sample. By fitting the linewidth of the lower-branch guided-mode resonance, a linewidth narrowing is observed near $k_x = 0$ μm$^{-1}$, confirming the formation of the BIC, shown in Fig. S2.

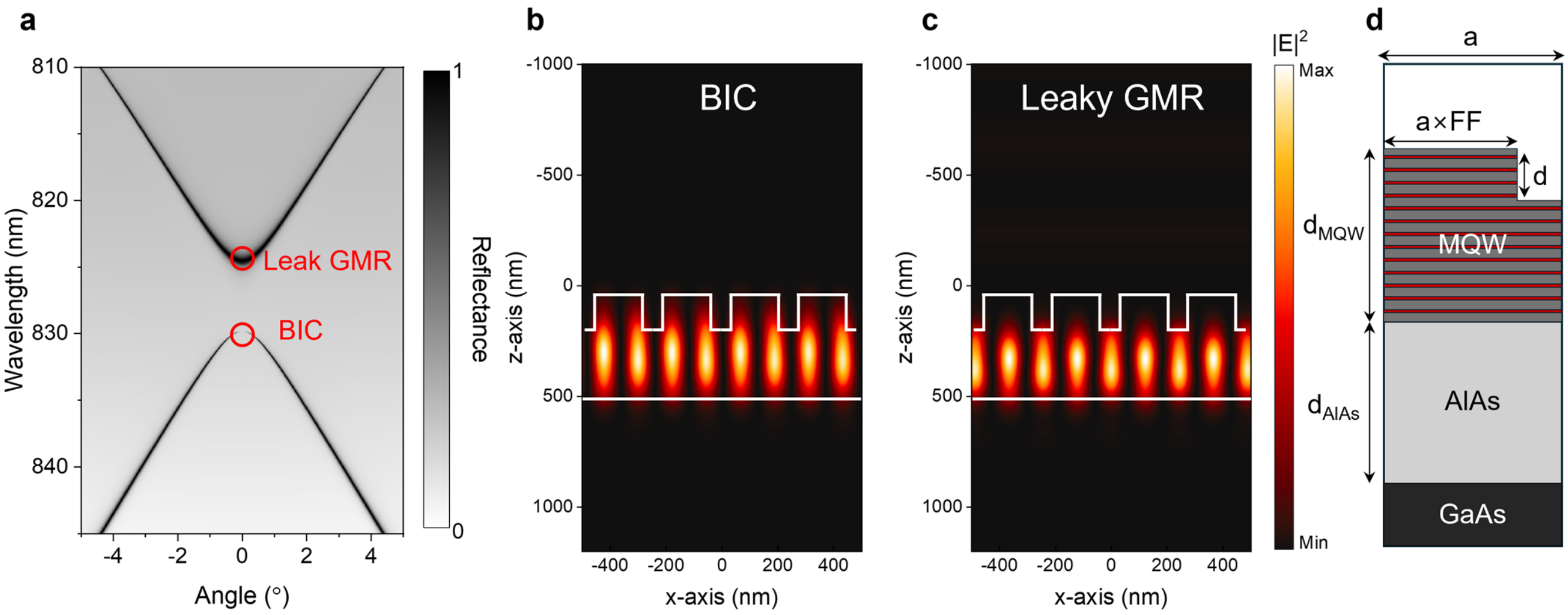

**Fig. S1. Simulation of guided mode resonances from 1D photonic crystal slab. a**, RCWA-calculated angle-resolved spectral map showing the photonic bands associated with the symmetry-protected BIC and the leaky GMR. **b, c**, Calculated electric-field intensity distributions, for the BIC and leaky GMR, respectively. Since the BIC is not accessible by excitation in RCWA, the field distribution is calculated by a plane wave with a small but finite incidence angle away from the 0°. The white outlines indicate the grating profile. **d**, Cross-sectional schematic of the unit cell. FF denotes the filling factor.

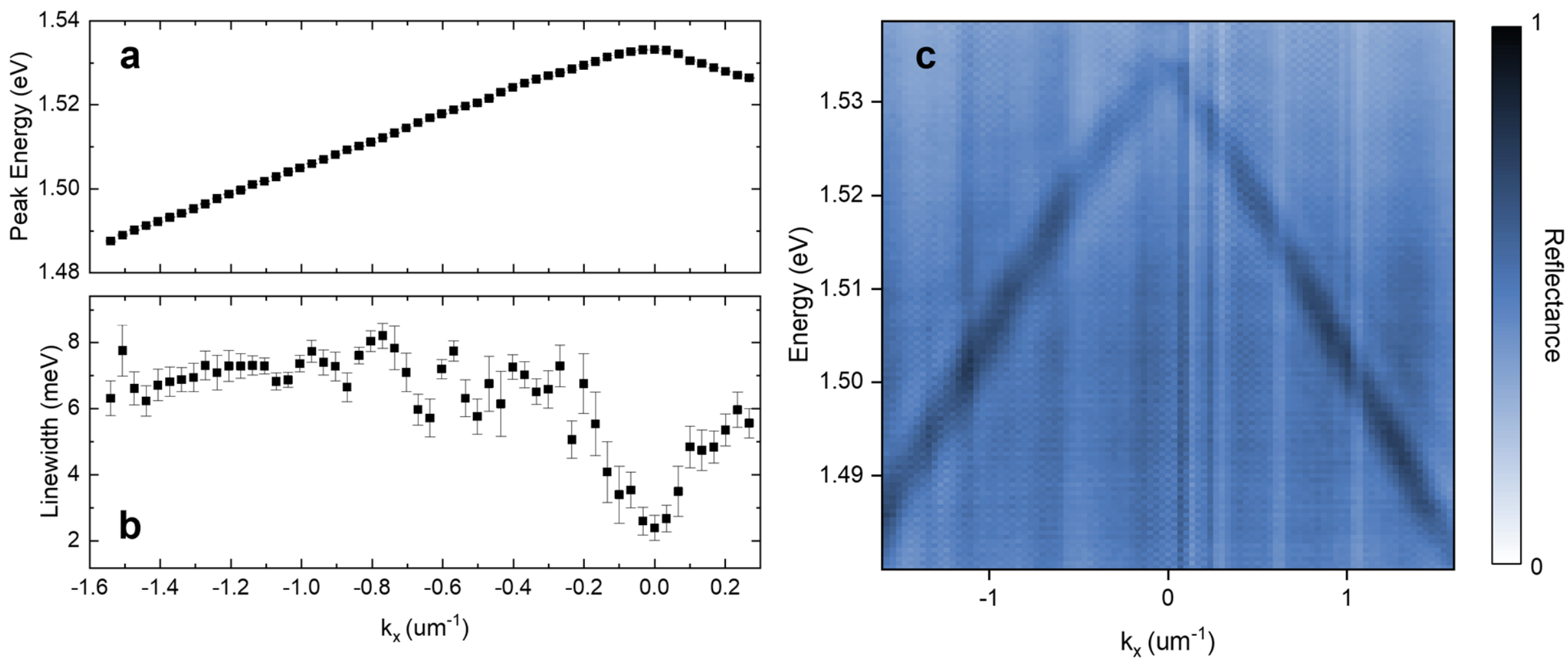

**Fig. S2. Experimental confirmation of a bound state in the continuum. a**, Resonance energy extracted from the angle-resolved reflectance spectrum as a function of the in-plane wavevector $k_x$. **b**, Resonance linewidth, showing linewidth narrowing toward the $k_x = 0\ \mu m^{-1}$. **c,** Angle-resolved reflectance spectrum of a uniform lattice with a = 244 nm.

## S2. EXCITON-POLARITONS FROM A 1D PHOTONIC CRYSTAL SLAB

The guided-mode resonance of the photonic crystal strongly couples to the excitonic resonance of the GaAs/$Al_{0.15}Ga_{0.85}As$ multi-quantum well. Because the guided-mode resonance and the exciton have different dispersions and effective masses, the detuning between them varies with wavevector. This strong-coupling behavior is evident in the dispersion measured by angle-resolved reflectance on uniform-lattice samples. Dispersions for both transverse-electric (TE, Fig. S3) and transverse-magnetic (TM, Fig. S4) polarizations show anti-crossing near zero detuning, indicating strong coupling between the exciton and the guided-mode resonance. Fitting the dispersion with a coupled-oscillator model yields a Rabi splitting of 15 meV for the TE polarization (Fig. S5).

Two excitonic resonances are observed in the spectra, the light-hole (LH) exciton at E = 1.564 eV and the heavy-hole (HH) exciton at E = 1.553 eV. From the optical selection rule, TE–HH and TM–LH coupling dominate, TE–LH coupling is one-third as strong as TE–HH, and TM–HH coupling is forbidden [2]. This behavior is consistent with the distinct polarization-dependent coupling observed in the angle-resolved reflectance measurements.

As the lattice constant increases, the polariton energy can be tuned according to the second-order Bragg condition. From the angle-resolved reflectance measurements, excluding strong-coupling effects, we find that the photonic mode energy shifts by 5 meV/nm as a function of lattice constant.

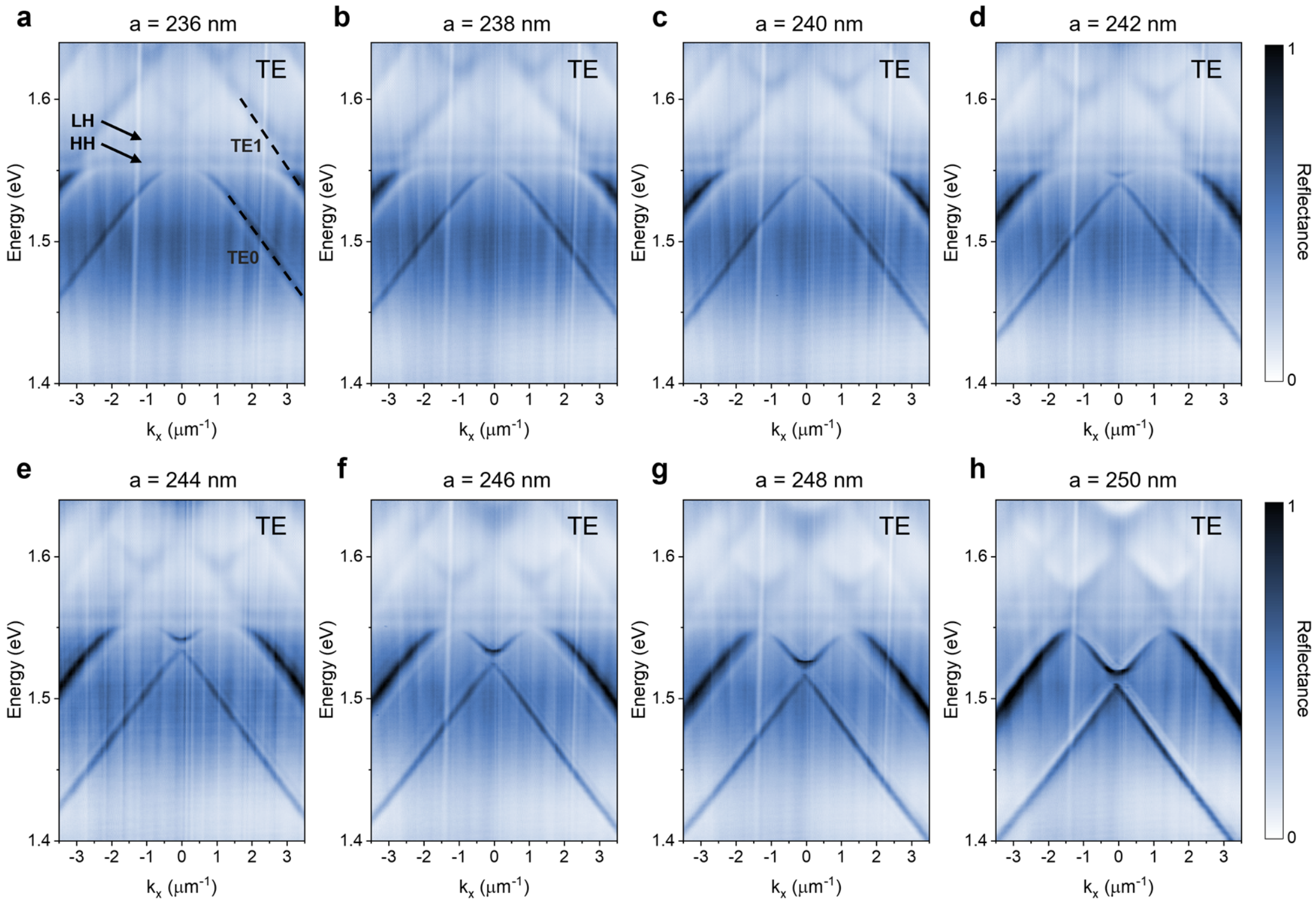

**Fig. S3. Polariton dispersion of TE polarization. a-h,** TE-polarized angle-resolved reflectance spectra of 1D photonic-crystal slabs with different lattice constants. The HH and LH exciton resonances are indicated by arrows in **a**. Dashed lines indicate TE0 and TE1 GMRs.

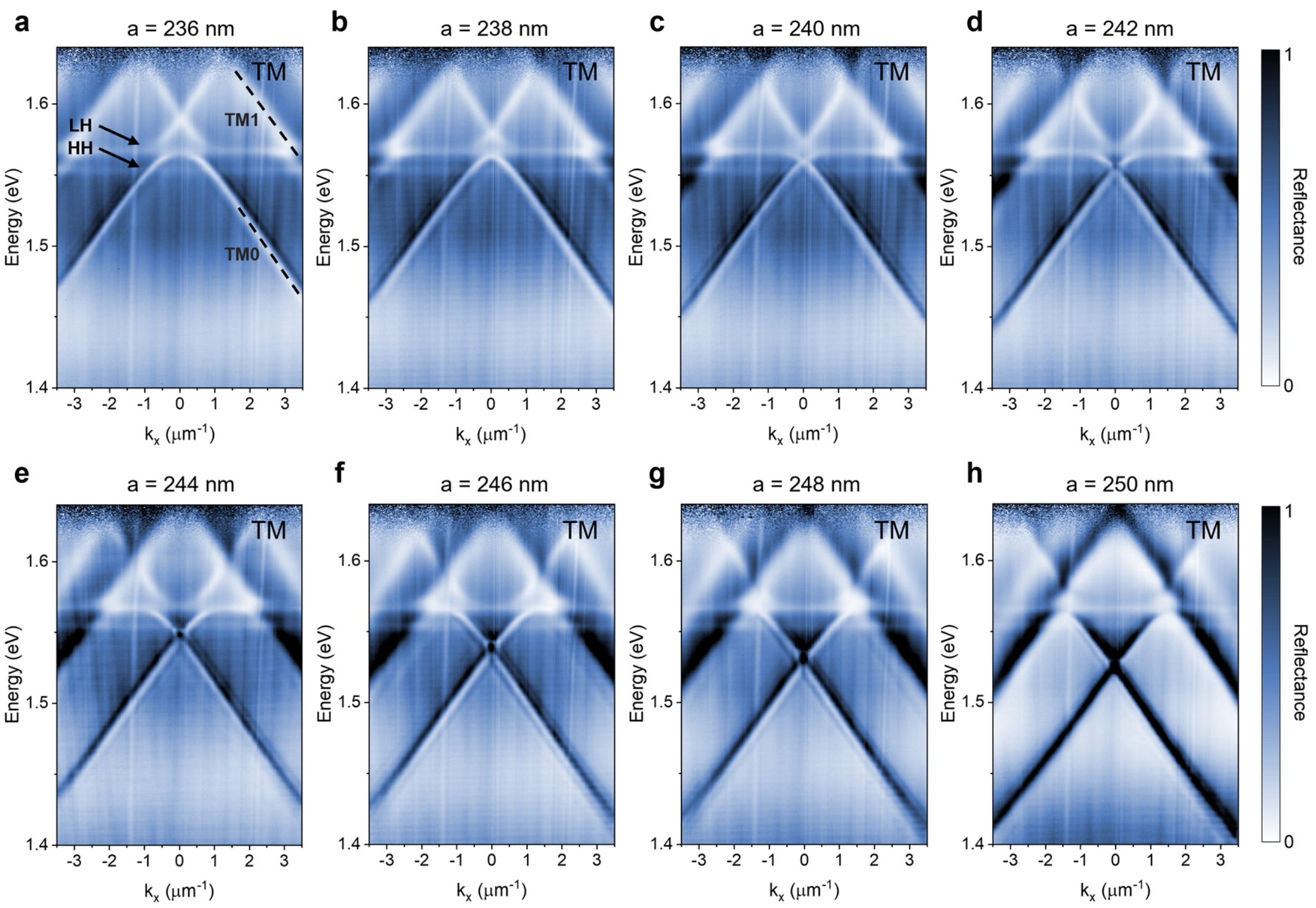

**Fig. S4. Polariton dispersion of TM polarization. a-h,** TM-polarized angle-resolved reflectance spectra of 1D photonic-crystal slabs with different lattice constants. The HH and LH exciton resonances are indicated by arrows in **a**. Dashed lines indicate TM0 and TM1 GMRs.

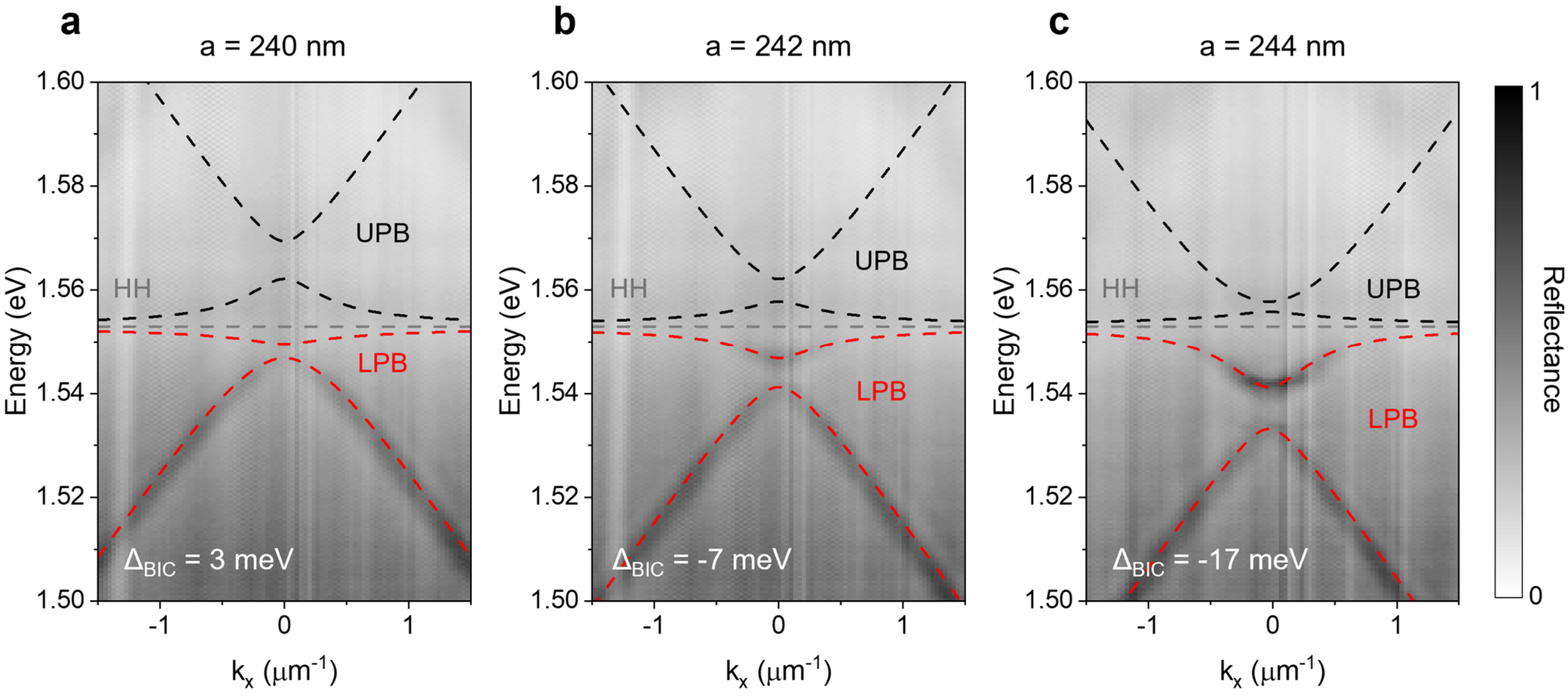

**Fig. S5. Strong coupling of TE-polarized guided mode resonance and heavy hole exciton resonance. a-c,** Angle resolved reflectance spectra for different lattice constant of a = 240, 242, and 244 nm. Dashed lines indicate the fitted dispersion from the coupled oscillator model. HH denotes the heavy-hole exciton resonance, and UPB and LPB denote the upper and lower polariton branches, respectively. $\Delta_{BIC}$ denotes the detuning of the bare BIC mode from the HH exciton resonance, defined as $\Delta_{BIC} = E_{BIC} - E_{HH}$.

## S3. TRAPPING EXCITON-POLARITONS WITH A LATTICE CONSTANT MOULATION

As described in the main text, the polariton potential arises from spatially modulating the lattice constant. For topological switching, we utilize a single trap formed as a photonic crystal well, shown in the scanning electron microscope image in Fig. S6. The angle-resolved PL spectra in the main text clearly show the trapped polariton. To further confirm that this polariton potential originates from the lattice constant modulation rather than excitation related effects, we performed angle-resolved reflectance measurements, showing that the trapped polariton state arises even without excitation (Fig. S7). Photonic confinement is simulated using the finite-difference time-domain (FDTD) method implemented in Ansys Lumerical 3D electromagnetic solver, shown in Fig. S8. The photonic dispersion is obtained by injecting a plane wave at varying incidence angles and capturing the reflectance. For a confined state, the spatial field distribution is extracted, showing clear confinement within the well region and decay into the barrier region.

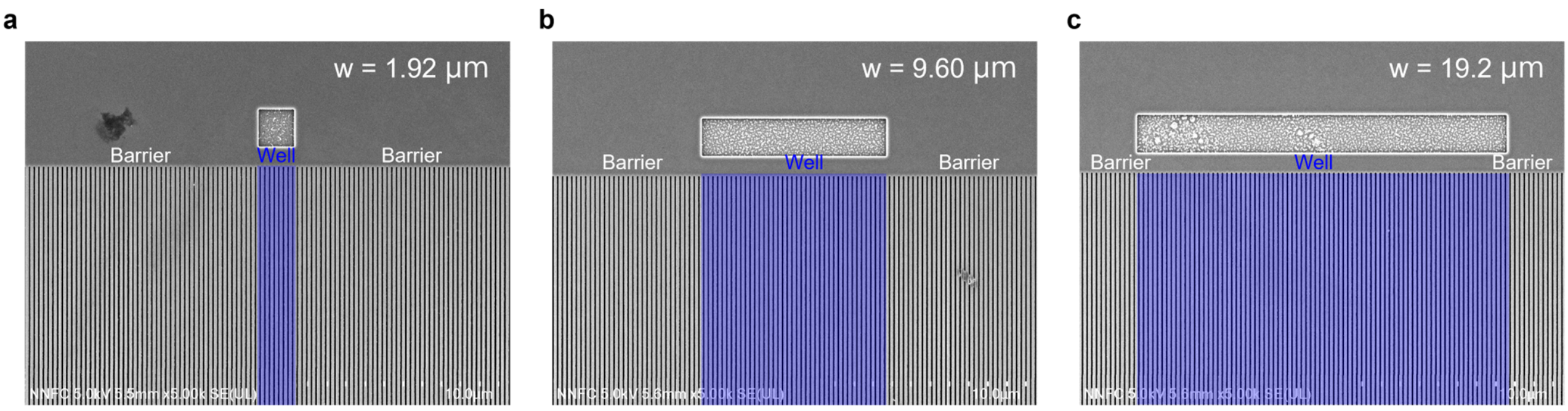

**Fig. S6. Scanning electron microscopy images of photonic crystal wells. a-c,** Top-view images of photonic crystal well structures with well widths of $w$ = 1.92, 9.60, and 19.2 μm, respectively. The well regions are highlighted in blue and are laterally confined by the surrounding photonic crystal barriers.

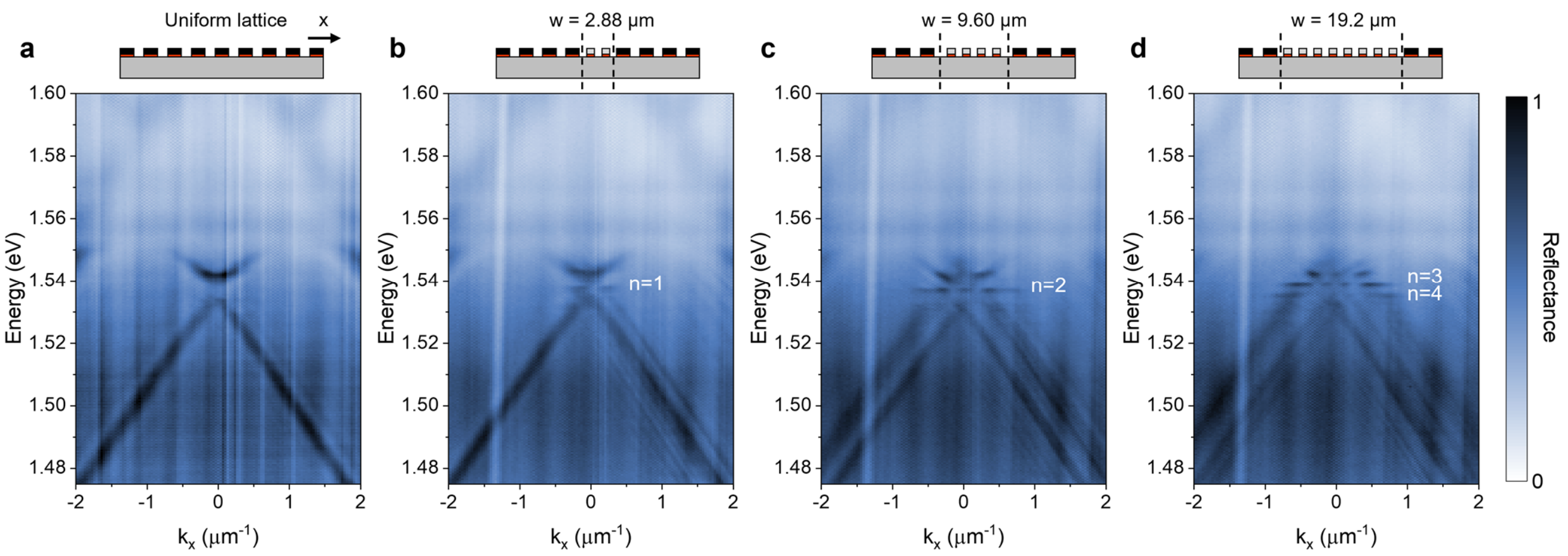

**Fig. S7. Dispersion of trapped polaritons from angle-resolved reflectance measurement. a-d,** Angle-resolved reflectance spectra of uniform lattice and photonic crystal well structures with $w$ = 2.88, 9.60, and 19.2 μm, respectively. Confined states are observed in the samples with a well, indicating that potential trap is formed without the effect of excitation.

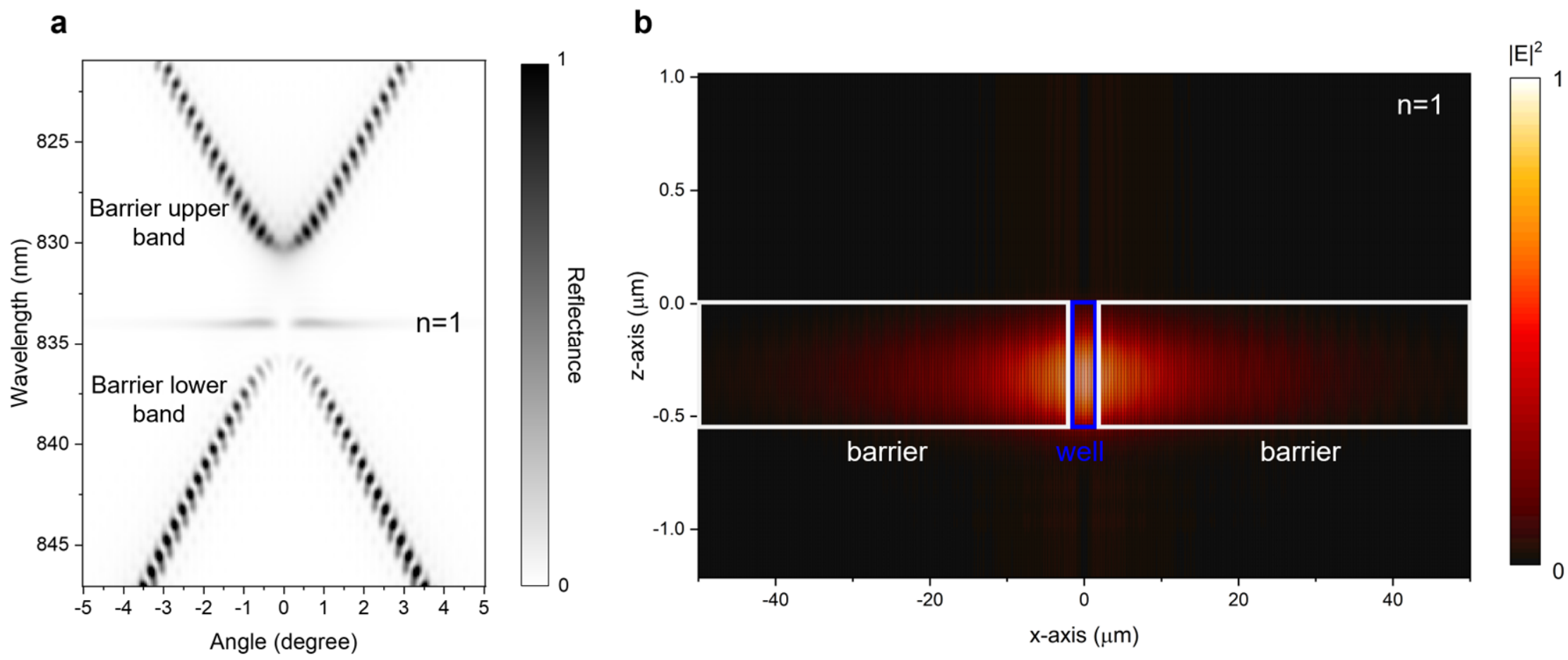

**Fig. S8. FDTD simulation of trapped photonic mode in a photonic crystal well. a**, Photonic dispersion of the photonic crystal well structure from FDTD simulation. **b**, Electric field intensity distribution of the $n$ = 1 mode. The mode is confined in the well region and decays into the barrier region.

## S4. ANALYTIC SOLUTION FOR MASSIVE DIRAC EQUANTION WITH A FINITE POTENTIAL WELL

To calculate the analytic solution for polariton confinement in the photonic crystal, we simplify the governing equation to a 1D Dirac equation by neglecting the strong-coupling term and the loss term from Eq. (1) in the main text.

$$\left(\frac{\hbar}{i}v\sigma_x\frac{\partial}{\partial x}+U\sigma_z+V(x)\right)\psi=E\psi \tag{S1}$$

Here, the polariton potential $V$(x) is set as a square well extending from x = -L/2 to x = L/2, with height $V_0$. To confine the negative-mass branch, the potential well is instead defined as a potential hill. With a spinor wave function $\psi=\begin{pmatrix}\psi_1\\ \psi_2\end{pmatrix}$, Eq. (S1) can be expanded into the following coupled first-order differential equations:

$$\frac{d}{dx}\psi_1=\frac{i}{\hbar v}[\{E-V(x)\}\psi_2+U\psi_2] \tag{S2}$$
$$\frac{d}{dx}\psi_2=\frac{i}{\hbar v}[\{E-V(x)\}\psi_1-U\psi_1] \tag{S3}$$

With continuous boundary conditions on $\psi_1$ and $\psi_2$, the eigenvalue equations for the even and odd eigenstate solutions can be derived as follows, where even and odd refer to the parity of $\psi_1$:

$$\tan(kL/2)=-\frac{E-U}{E-V_0-U}\frac{k}{\kappa} \tag{S4}$$
$$\cot(kL/2)=\frac{E-U}{E-V_0-U}\frac{k}{\kappa} \tag{S5}$$

Here, the evanescent decay constant $\kappa$ is given by $\kappa=\frac{1}{\hbar v}\sqrt{U^2-E^2}$, and wave number $k$ is given by $k=\frac{1}{\hbar v}\sqrt{(E-V_0)^2-U^2}$. Solving these equations yields the theoretically calculated bound state eigenenergies presented in the main text. Note that when the eigenenergy exceeds the barrier bandgap ($|E|>|U|$), the decay constant $\kappa$ becomes imaginary, meaning the state no longer remains bound inside the well and instead propagates through the barrier region.

## S5. TRAPPED POLARITON CONDENSATES WITH WELL WIDTH VARIATION

In the uniform lattice, a slight potential arising from exciton reservoirs on an excitation spot confines polaritons [3]. This leads to condensation into the $n = 1$ state, shown in Fig. S9. Above the condensation threshold, the PL intensity increases super linearly and linewidth narrowing is observed. With a small trap structure of $w = 1.92$ µm, polaritons condense into the confined $n = 1$ state, as described in Fig. 3 of the main text. With increasing well width, the $n = 1$ state no longer lies within the barrier bandgap, and other confined states, such as $n = 2$ or $n = 3$, condense instead, depending on the well width (Fig. S10). Spatially resolved PL spectra of the $n = 2$ and $n = 3$ states show confinement within the well region (Fig. S11). Each confined state acquires an additional spatial node when measured in the far field [4].

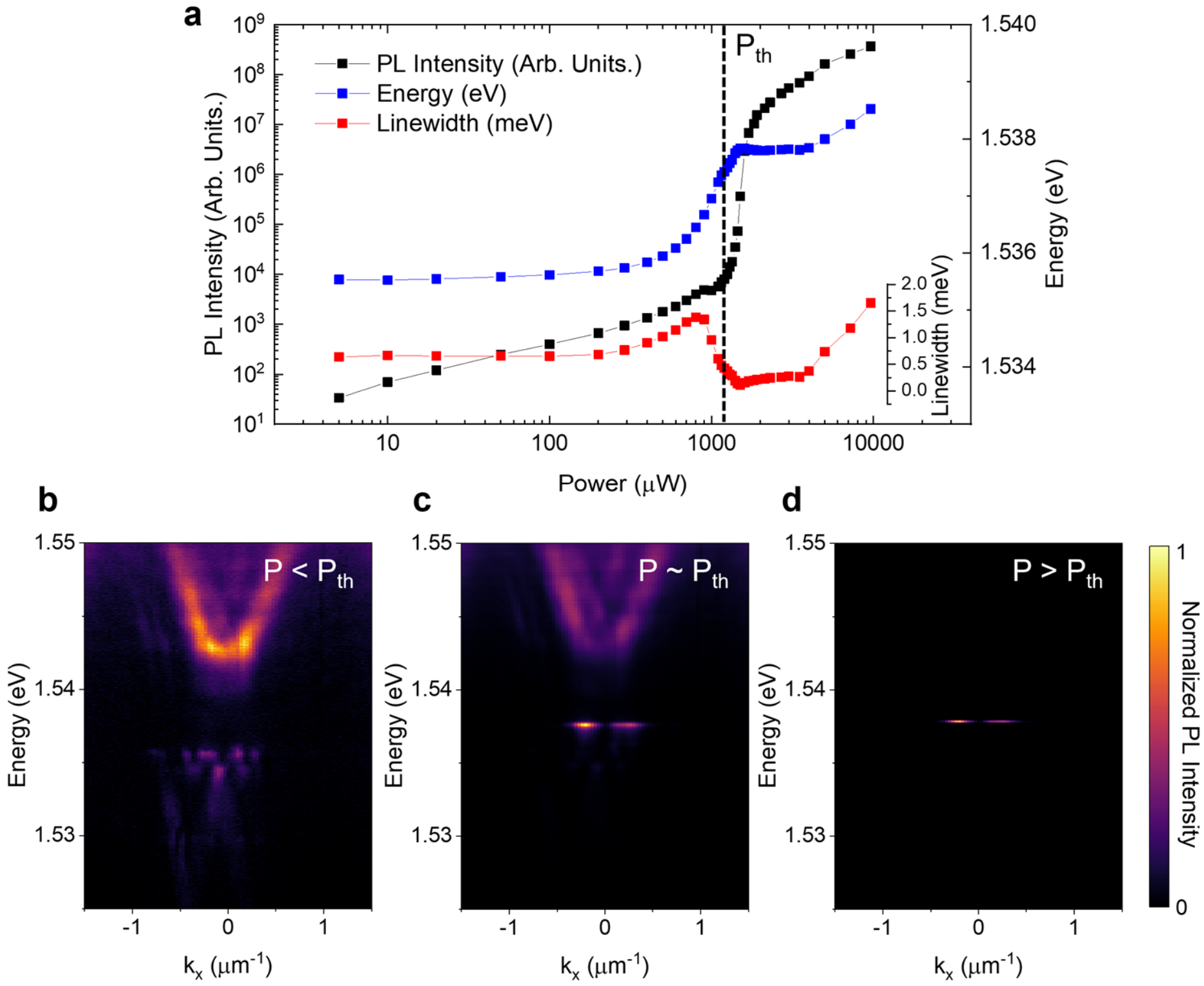

**Fig. S9. Polariton condensation in a uniform lattice. a**, Excitation power dependent PL intensity, energy, and linewidth of the polariton in a uniform lattice. **b-d**, polariton dispersion for different excitation powers, captured by angle-resolved PL measurement.

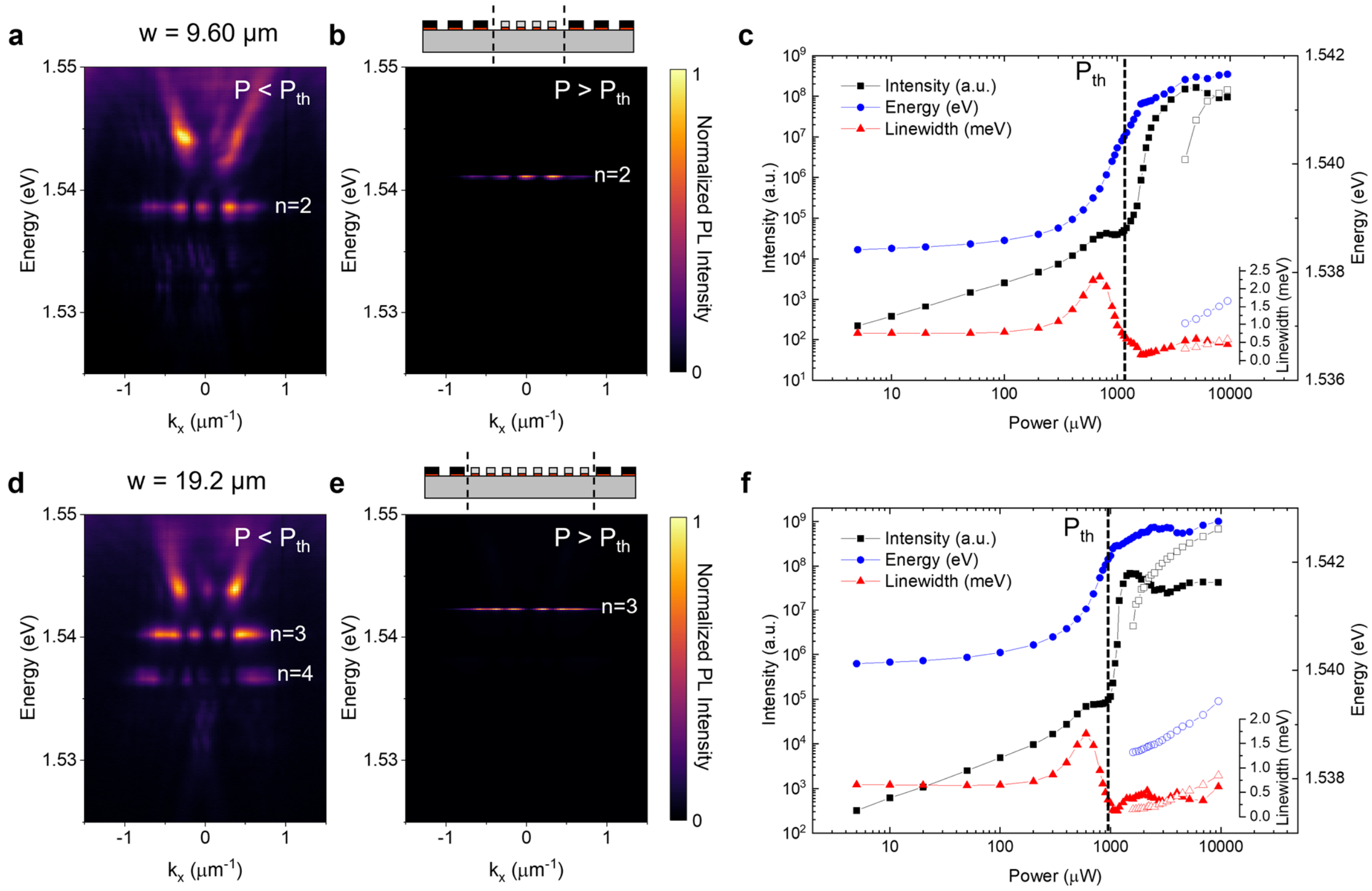

**Fig. S10. Trapped polariton condensation with various well width. a, b**, Angle-resolved PL spectra for $w = 9.60$ μm with excitation power before and after condensation threshold. Polaritons condense in the $n = 2$ state. **c**, Excitation power dependent PL intensity, energy, and linewidth of the polariton for $w = 9.60$ μm well. open symbol indicates the $n = 3$ state. **d, e**, Angle-resolved PL spectra for $w = 19.2$ μm with excitation power before and after condensation threshold. Polaritons condense in the $n = 3$ state. **f**, Excitation power dependent PL intensity, energy, and linewidth of the polariton for $w = 19.2$ μm well. open symbol indicates the $n = 4$ state.

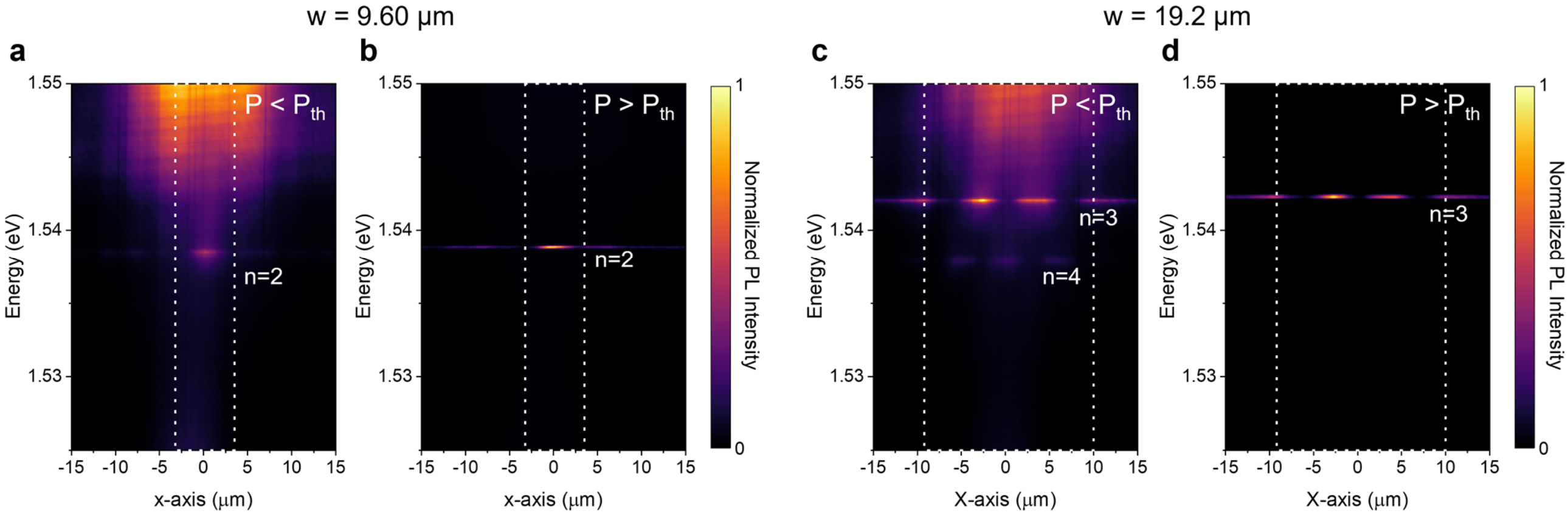

**Fig. S11. Spatial distribution of trapped polariton condensates. a, b**, Spatial resolved PL spectra for $w = 9.60$ μm with excitation power before and after threshold, respectively. **c, d**, Spatial resolved PL spectra for $w = 19.2$ μm with excitation power before and after threshold, respectively. White dashed box indicates the well region.

## S6. SPATIAL PHASE DISTRIBUTION OF TRAPPED POALRITON CONDENSATES

After polariton condensation, the established coherence allows us to interfere the emitted photons from the condensate and extract the phase map. To achieve this, we use a Mach-Zehnder interferometer to interfere the signal with a phase reference. By adjusting the optical delay of the signal arm with a motorized stage within the temporal coherence time, an interference image is constructed on a charge-coupled device (CCD) camera. To separate the sidebands from the main peak in the Fourier-transformed interferogram, a tilted reference beam with carrier frequency $k_c$ is used. At the CCD camera plane, the electric fields of the signal and reference arms are given by the following equations:

$$E_s = A_s(\boldsymbol{r})e^{i\Phi_s(\boldsymbol{r})} \tag{S6}$$

$$E_r = A_r(\boldsymbol{r})e^{i\{\Phi_r(\boldsymbol{r})+\boldsymbol{k_c}\cdot\boldsymbol{r}\}} \tag{S7}$$

We assume the phase of the reference arm is constant. The measured interference intensity is given by:

$$I_{CCD}(\boldsymbol{r}) = I_s(\boldsymbol{r}) + I_r(\boldsymbol{r}) + E_s(\boldsymbol{r})E_r^*(\boldsymbol{r}) + E_r(\boldsymbol{r})E_s^*(\boldsymbol{r}) \tag{S8}$$

The measured real space interference image is subtracted by the signal and reference images, then Fourier transformed into reciprocal space. The Fourier transformed image is given by:

$$\tilde{I}(\boldsymbol{k}) - \tilde{I}_s - \tilde{I}_r = \mathcal{F}\{A_sA_re^{i\{\Phi_s(\boldsymbol{r})-\Phi_r\}}\}(\boldsymbol{k}-\boldsymbol{k_c}) + \mathcal{F}\{A_sA_re^{i\{\Phi_s(\boldsymbol{r})-\Phi_r\}}\}(\boldsymbol{k}+\boldsymbol{k_c}) \tag{S9}$$

By filtering the Fourier transformed image near the carrier frequency and transforming it back to real space, we obtain the phase by taking the argument of the filtered field.

The phase map of the ground state polariton condensate shows a single π-phase jump across the antinode, as seen in Fig. S12 [5]. Higher confined mode condensates for different well widths show a number of π-phase jumps corresponding to the confinement quantum number $n$.

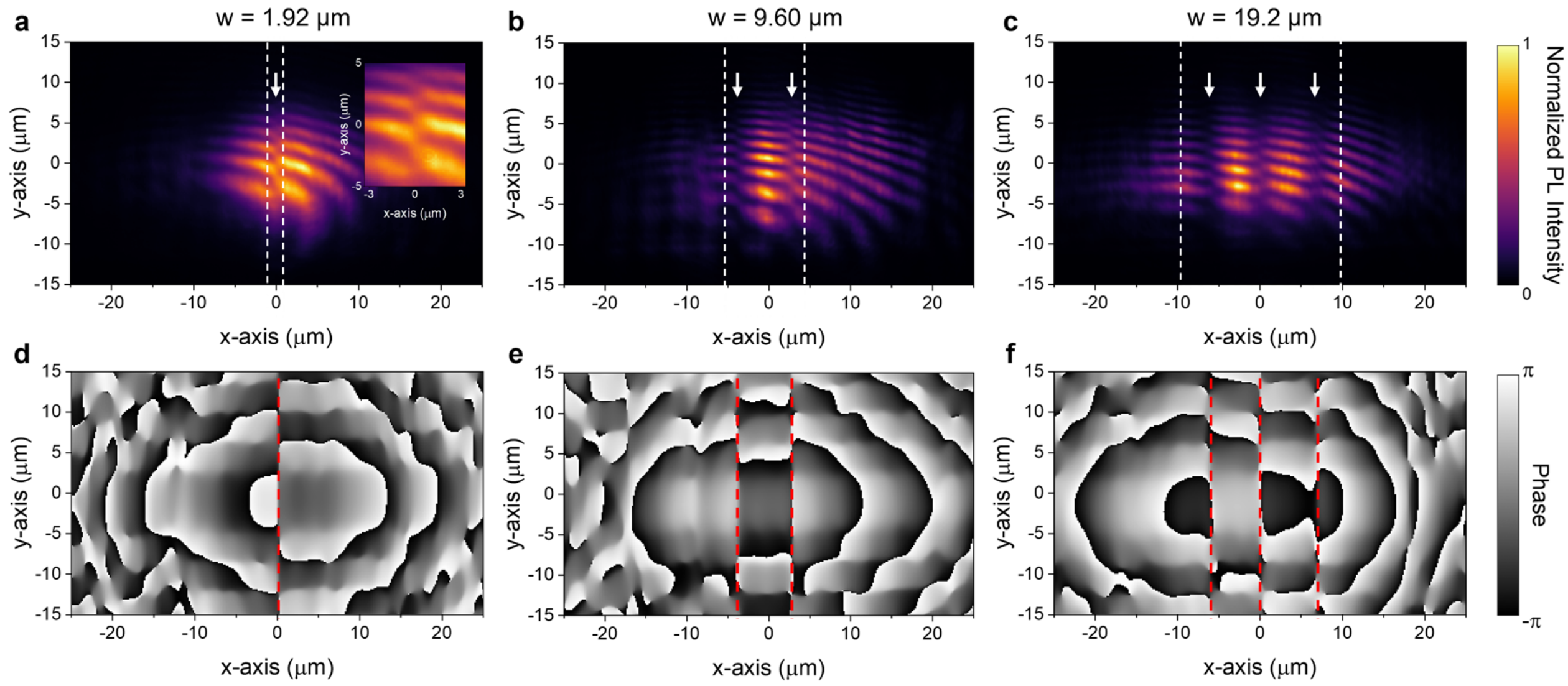

**Fig. S12. Spatial phase distribution of trapped polariton condensates. a-c**, Interferometric PL images of condensates of $n$ = 1, $n$ = 2 and $n$ = 3 states in $w$ = 1.92, 9.60, and 19.2 μm wells, respectively. The inset in **a** shows magnified image for clarity. White arrows indicate spatial nodes with a phase jump and white dashed lines indicate the well region. **d-f**, Reconstructed spatial phase distributions of trapped polariton condensates. Red dashed lines indicate a phase jump.

## S7. BLUESHIFT OF TRAPPED STATE AND BARRIER MODE POLARITONS

To demonstrate switching between confined topological polariton condensates, the confined state needs to blueshift into the barrier band and leak out into the barrier with increasing excitation power. The prerequisite for this scheme is that the barrier energy experiences a smaller blueshift compared to the confined mode. Since the excitation beam with a Gaussian beam spot of 7.6 µm full width at half maximum is centered within the well region, most of the blueshift is expected to occur in the confined state. Additionally, polariton confinement further enhances the blueshift relative to barrier band polaritons. To experimentally examine this, the blueshift of the confined state is compared with that of the two barrier band edges below the condensation threshold. The confined $n = 1$ state shows a clear blueshift relative to the two barrier band edge states, as described in Fig. S13. Above the condensation threshold at $P_{th}$ = 685 µW, the barrier band edge states became suppressed relative to the $n = 1$ or $n = 2$ states, precluding reliable tracking. However, we expect only a marginal blueshift of the barrier band edge states above the condensation threshold compared with those below the threshold, owing to the reduced exciton reservoir density.

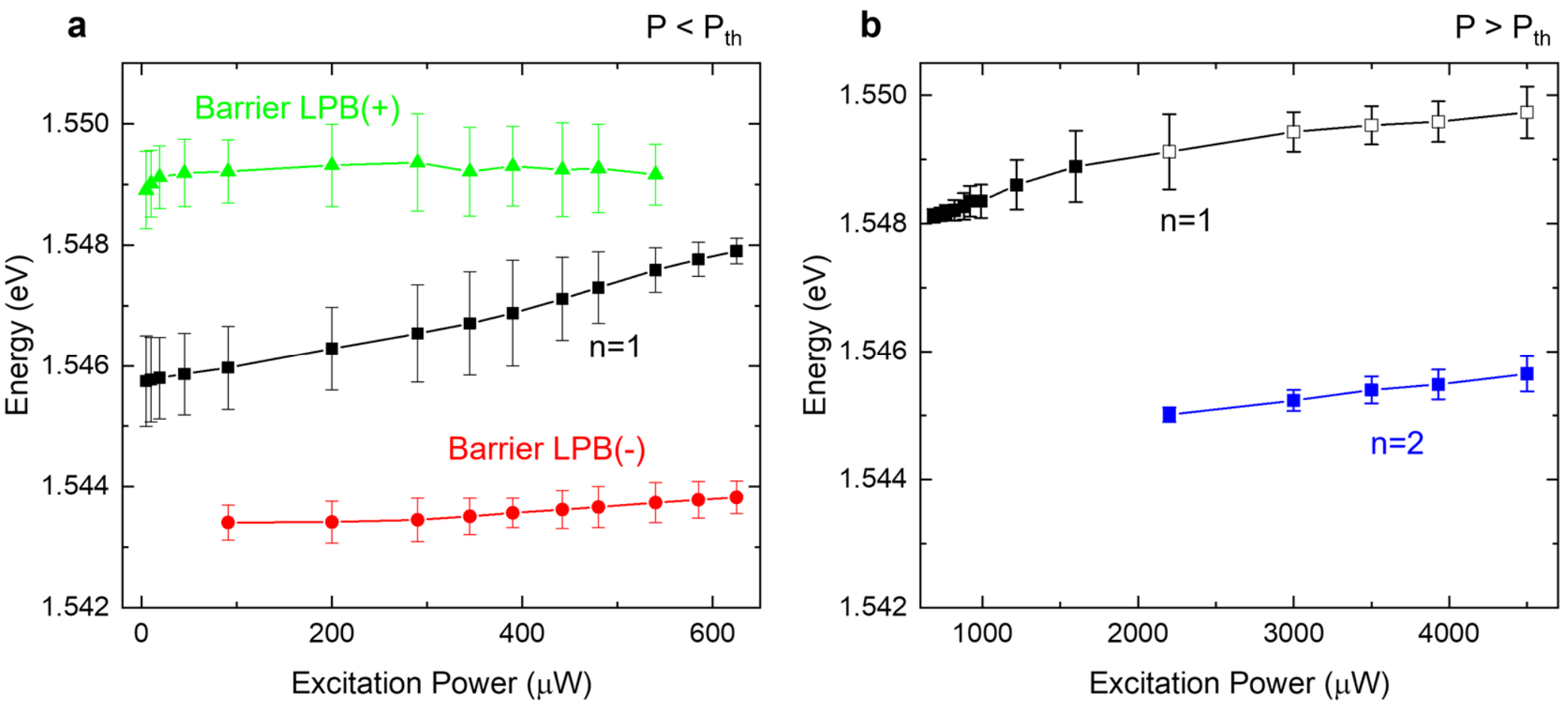

**Fig. S13. Comparison of blueshift of tapped polaritons and barrier polaritons. a.** Polariton blueshift behavior captured below the condensation threshold for the confined polariton sample with well width of $w$ = 2.86 µm. Error bars indicate the linewidth of each mode. The $n = 1$ state blueshifts dominantly, since its overlap with the excitation spot is large. This provides a pathway to tune the trapped state independently of the barrier, giving a knob to actively tune polariton condensates. **b.** Polariton blueshift behavior above the condensation threshold. Above the condensation threshold, barrier band edge mode population is suppressed. Filled black squares indicate the $n = 1$ state energy before the $n = 2$ state (filled blue diamonds) becomes dominant, and open black squares indicate the energy of suppressed $n = 1$ state after the $n = 2$ state becomes dominant. As the $n = 2$ state emerges, the blueshift slope of the $n = 1$ state exhibits a decrease, which can be attributed to a reduction in polariton density within the $n = 1$ state.

## S8. POLARIZATION VORTEX OF POLARITON CONDENSATION

Since carriers formed by non-resonant excitation relax into the lowest-energy exciton reservoir, the HH exciton and its related polariton become dominant in the PL measurement. Due to this selectivity, the TE-polarized guided mode resonance captures most of the polariton population and dominates the emission. Thus, TM-polarized polaritons can be ignored at any angle of incidence, and only the BIC mode from the TE-polarized guided mode resonance is considered in the main text and in what follows.

The polarization vortex in k-space is reconstructed by calculating the Stokes parameters from the polarization-resolved angle-resolved images using the following equations:

$$\varphi = \frac{1}{2}\tan^{-1}\left(\frac{S_2}{S_1}\right) \tag{S9}$$

where the Stokes parameters $S_1 = \frac{I_H - I_V}{I_H + I_V}$ and $S_2 = \frac{I_D - I_A}{I_D + I_A}$ are calculated in each k-space point. Based on this calculation, the polarization vortex of the trapped polariton condensate is shown in the main text. As a comparison, the polarization vortex of the BIC polariton condensate in the uniform lattice shows a topological charge of $q = +1$, shown in Fig. S14.

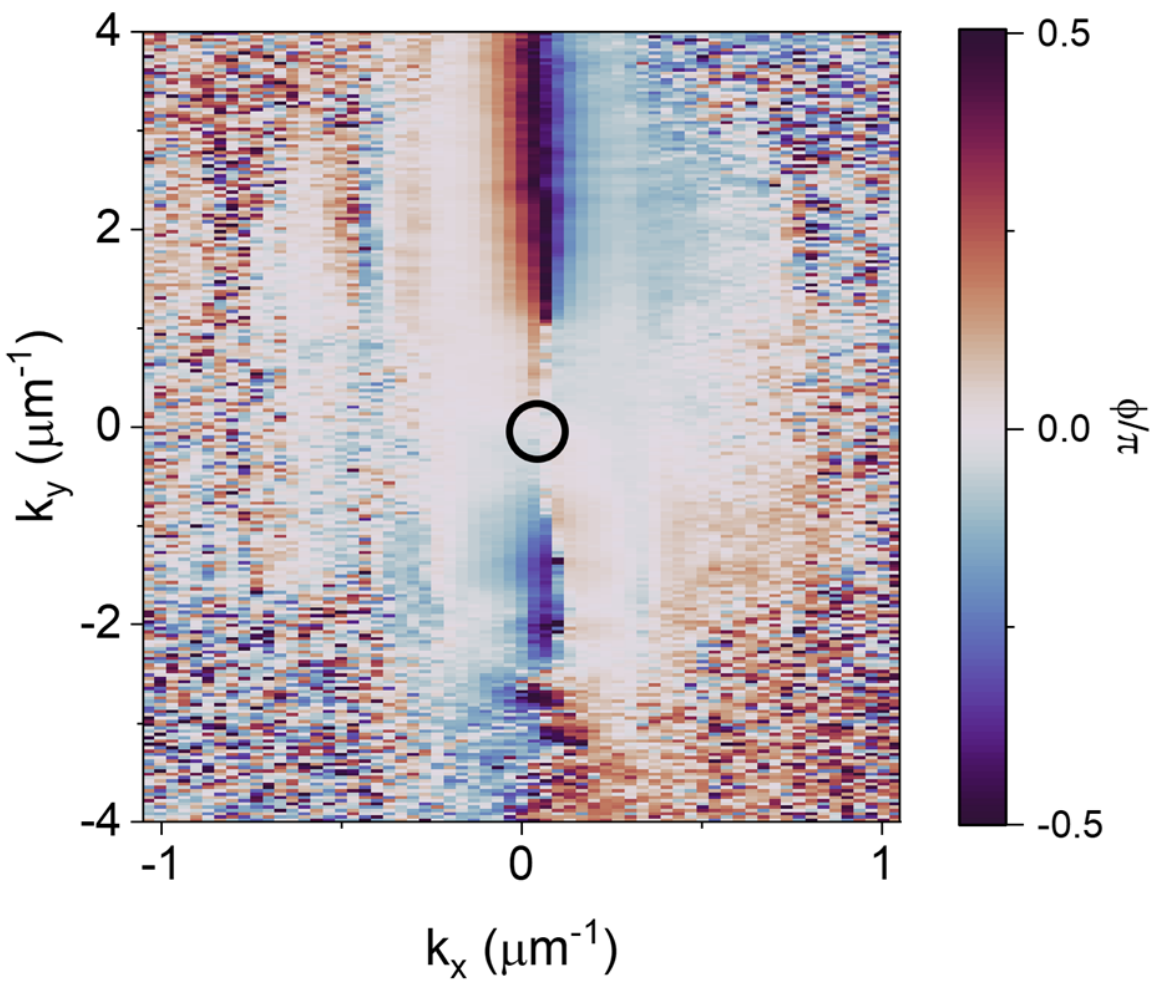

**Fig. S14. Polarization vortex of polariton condensates in a uniform lattice.** Polarization angle map in k-space for polariton condensates in a uniform lattice. A polarization vortex with a topological charge of $q = +1$ is observed.